\documentclass[sigconf, nonacm]{acmart}

\AtBeginDocument{%
  \providecommand\BibTeX{{%
    \normalfont B\kern-0.5em{\scshape i\kern-0.25em b}\kern-0.8em\TeX}}}

\setcopyright{acmcopyright}
\copyrightyear{2024}
\acmYear{2024}
\acmDOI{XXXXXXX.XXXXXXX}
\usepackage{booktabs}

\usepackage{comment,enumitem,setspace,float}

\begin{document}

\title{The MovieLens Beliefs Dataset: Collecting Pre-Choice Data for Online Recommender Systems}
\author{Guy Aridor}
\email{guy.aridor@kellogg.northwestern.edu}
\affiliation{%
  \institution{Northwestern Kellogg}
  \country{United States}
}

\author{Duarte Gon\c{c}alves}
\email{duarte.goncalves@ucl.ac.uk}
\affiliation{%
  \institution{University College London}
  \country{United Kingdom}
}

\author{Ruoyan Kong}
\email{kong0135@umn.edu}
\affiliation{%
  \institution{University of Minnesota - Twin Cities}
  \country{United States}
}

\author{Daniel Kluver}
\email{kluve018@umn.edu}
\affiliation{%
 \institution{University of Minnesota - Twin Cities}
 \country{United States}}

\author{Joseph A. Konstan}
\email{konstan@umn.edu}
\affiliation{%
  \institution{University of Minnesota - Twin Cities}
  \country{United States}}

\renewcommand{\shortauthors}{Aridor et al.}


\begin{abstract}
    An increasingly important aspect of designing recommender systems involves considering how recommendations will influence consumer choices. 
    This paper addresses this issue by introducing a method for collecting user beliefs about un-experienced goods – a critical predictor of choice behavior. 
    We implemented this method on the MovieLens platform, resulting in a rich dataset that combines user ratings, beliefs, and observed recommendations. 
    We document challenges to such data collection, including selection bias in response and limited coverage of the product space. 
    This unique resource empowers researchers to delve deeper into user behavior and analyze user choices absent recommendations, measure the effectiveness of recommendations, and prototype algorithms that leverage user belief data, ultimately leading to more impactful recommender systems. The dataset can be found at \url{https://grouplens.org/datasets/movielens/ml_belief_2024/}.
\end{abstract}

\maketitle

\section{Introduction}

The overwhelming abundance of goods on online platforms creates a challenging environment for consumers to find the best alternatives for them.
Recommender systems aim to bridge this gap by suggesting goods to consumers.
Traditionally, these systems focused on predicting how much user $i$ will value each good $j$ (denoted $u_{i,j}$)  \citep{adomavicius2005toward}.
In order to do so, platforms can collect a wide range of data, including explicit user inputs of ratings or other on-platform behavior, such as past consumption and search time.

Early on in recommender system research it was recognized that accurately predicting user's consumption value $u_{i,j}$ might not be sufficient for recommendations to be useful for users \citep{mcnee2006being}.
Since then a large literature has arisen that not only utilizes predicted consumption value but also incorporates auxiliary criteria in recommendations \citep{kaminskas2016diversity}, giving rise to approaches based on serendipity \citep{ge2010beyond, kotkov2016challenges, kotkov2018investigating}, novelty \citep{vargas2011rank, castells2021novelty}, and calibration \citep{steck2018calibrated, abdollahpouri2020connection}, among others. These different approaches arise from the intuition that useful recommendations have to take into account how they can be useful in the user decision-making process. 
However, they neither explicitly consider the user choice problem nor specify through which mechanisms recommendations can assist users.

Previous research has shown that one useful type of data for understanding user choice and the mechanisms through which recommendations act is user \textit{belief} data --- the opinions that users have about goods they \textit{have not} consumed. 
\cite{AridorGoncalvesSikdar2020ACMRecSys} show through numerical simulations of a theoretical user choice model that user beliefs can rationalize empirical consumption patterns documented by \cite{NguyenHuiHarperTerveenKonstan2014ACM} both with and without recommendations on MovieLens. 
Furthermore, \cite{aridor2023economics} conduct a field experiment on MovieLens and show that recommendations have a causal effect on user beliefs and that user beliefs predict what they will consume.

\noindent \textbf{Our Contributions}: In this paper we provide both a procedure to effectively collect belief data and an open-source dataset generated by implementing the procedure on the MovieLens platform for over a year. 
We rely on a simple economic model of decision-making in the context of recommender systems to guide the type of variation that is generated by our procedure. 
In particular, we will showcase how the procedure generates data that makes it feasible to predict beliefs over the full set of goods, measure how recommendations influence beliefs, and predict how beliefs map to consumption. 

While our procedure is motivated by a particular model of decision-making, there are many possible applications of the procedure and the data. 
The procedure is designed so that it can be implemented and scaled on any online platform that deploys a recommender system. In particular, the procedure carefully chooses the set of feasible goods and how to pick which goods from this feasible set to elicit beliefs about in a manner that overcomes endemic challenges to collecting data regarding non-consumption behavior on online platforms. The resulting open-source data complements the widely popular MovieLens ratings dataset \citep{HarperKonstan2015ACMTIIS} and can be useful for the canonical ratings prediction problem. Existing research exploits the fact that ratings are missing not at random either as a source of information \citep{ying2006leveraging} or in order to debias ratings \citep{schnabel2016recommendations}, but the belief data provides a distinct view of preferences that, for instance, can differentiate between parts of the product space where a user does not consume a good since she is uncertain about its quality versus because she knows that she does not like certain types of goods. 

As such, it provides a novel view into user preferences relative to ratings data that can also be used to test and design recommendation evaluation criteria that incorporate how recommendations influence user choices. For instance, one exciting possible application of the dataset is to use it to directly measure the degree of serendipity of a recommendation (e.g., ``unexpectedness" of a recommendation relative to user prior beliefs) and incorporate this into existing serendipity-based recommendation evaluation criteria. Beyond the several applications of the dataset we highlight throughout the paper, it contains a rich description of user behavior beyond the traditional MovieLens dataset --- data which we believe will prove useful for designing the next generation of recommender systems.

\section{Procedure}\label{sec:procedure}

\subsection{Economic Model of Decision-Making}\label{subsec:econ_model}

First, we document the basic model of user decision-making that guides the procedure, based on the proposals in \cite{AridorGoncalvesSikdar2020ACMRecSys, aridor2023economics}. 
We use the model primarily to guide what kind of data to collect as well as what variation we want to generate in our procedure.

We consider that there are $I$ consumers that make a sequence of choices from a choice set $X^t$, which are the baseline set of goods available at time $t$. 
For every good $n$, we assume that each user $i$ assigns a monetary equivalent $x_{i,n} \in \mathbb{R}$ to the experience of consuming it. 
Each user can value the same good differently. 
However, we assume that each user derives utility from money as given by a utility function $u_i: \mathbb{R} \rightarrow \mathbb{R}$, strictly increasing and continuous. 
In typical environments where recommender systems are deployed, these goods are experience goods and so users are uncertain about how much they will value each good. 
In particular, even users that will end up having the same ex-post valuation of good $n$ may differ in their ex-ante valuation because they hold different beliefs about it. 
We denote by $p_{i}$ the beliefs user $i$ has about how she will value each of the goods in the product space. 
Each user evaluates the good according to its expected utility, $\mathbb{E}_{p_{i}} [u_i(x_{i,n})]$.

Each time the consumer enters the platform, they receive a set of recommendations, denoted $r_{i,t}$. 
For now, we remain agnostic to how this set is generated. 
We consider that the recommendation directly shifts the user's beliefs so that the user's expected utility following the set of recommendations is given by $\mathbb{E}_{p_{i}} [u_i(x_{i,n}) \mid r_{i,t}]$. 
Thus, the user's choice in both scenarios is given by:
\begin{align*}
    c_{i}^{NREC} &:= \arg \max\limits_{n} \mathbb{E}_{p_{i}} [u_i(x_{i,n})], \\
    c_{i}^{REC}(r_{i,t}) &:= \arg \max\limits_{n} \mathbb{E}_{p_{i}} [u_i(x_{i,n}) \mid r_{i,t}].
\end{align*}

\noindent The consumer welfare-maximizing choice of $r_{i,t}$ for period $t$ is then
\begin{align*}
    r_{i,t}^{OPT} = \arg \max_{r_{i,t}} u_{i}^{*,REC}(r_{i,t})
\end{align*}
where $u_{i}^{*,REC}(r_{i,t}) := \max\limits_{n} \mathbb{E}_{p_{i}} [u_i(x_{i,n}) \mid r_{i,t}]$ denotes the maximized expected utility given the platform recommendation.

The intuition for this is fairly straightforward. 
The platform chooses a slate of recommendations that maximize the user's utility. 
Using this simple framework, it is also easy to rationalize the observation of \cite{mcnee2006being} that the most accurate recommendations are not always the most useful recommendations. 
This is because the value of recommendation does not come from an accurate prediction of $u_{i,n}$, but rather from information that shifts beliefs and subsequently consumption choices. 
For instance, recommendations have no value to consumers whatsoever if they are not effective in convincing the user to change their choices --- i.e. $c_i^{NREC}=c_{i}^{REC}(r_{i,t})$.
Moreover, one can view this as an economic formalism for serendipity as ``unexpected" and ``useful" recommendations can be specified as providing information on goods that the user will end up liking and that they did not expect to like before the information.

\raggedbottom

\subsection{The Data Collection Challenge}

The goal of the procedure is to select the set of movies to elicit beliefs about (denoted $M^t$) and how we sample from this set (denoted $B_i^t$). 
Guided by Section \ref{subsec:econ_model} we want to design the procedure so that we have the necessary variation to (1) characterize the set of prior beliefs over the space of goods, (2) predict how beliefs map to consumption, and (3) identify how recommendations shift beliefs.
A natural question is how this data collection exercise differs from the typical data collection for a recommender system. 
Unlike ratings or implicit consumption data, such data is not a natural byproduct of usage of the platform. 
For instance, on the MovieLens platform users will input their ratings for movies that they have seen and, in many cases, they actively seek out the movies to rate them. 
Furthermore, the distribution of movies that users consume follows a pattern that is only partially impacted by the platform's choices. 
This means that the data a platform has access to for designing good recommendations is unbalanced across different goods, potentially impacting the usefulness of recommendations.

An important difference is that the belief data has to be elicited instead of stemming from consumption activity directly.
Consequently, the platform needs to specify which parts of the product space the collection of this belief data should target for each user or user type. As such, to tackle challenge (1) we need to design the procedure in such a manner that we can get a representative enough sample of users and goods such that we can use the collected data to extrapolate across the full set of beliefs over the entirety of the product space. 
In principle, this is possible for the same reason that both collaborative filtering and content-based recommender systems work: there is correlation in ratings both across goods and across users. 
Indeed, \cite{aridor2023economics} document that this is the case for belief data collected on the MovieLens platform, which indicates that standard collaborative filtering and content-based methods may work to estimate the full set of beliefs. 
As a result, we design a procedure to select a subset of movies that we elicit beliefs about, $M^t$, which can then be extrapolated to the full set of movies. 
However, we face a tradeoff in the size of $M^t$: a larger $M^t$ gives us coverage of more movies but with potentially fewer beliefs per movie, whereas a smaller $M^t$ gives us less coverage but with a denser matrix of users $\times$ movies.
This impacts both the prediction problem of estimating the full set of beliefs (challenge (1)) and the exercise of predicting consumption from beliefs (challenge (2)).

Given sufficient data to characterize beliefs and to predict how beliefs map to consumption, we could predict consumption without recommendation, $c_{i}^{NREC}$. 
In order to know how recommendations would shift consumption behavior, we need to characterize how beliefs evolve with recommendations (i.e., $p_{i} \mid r_{i,t}$), which is precisely challenge (3). 
In our procedure, we do not modify the baseline set of recommendations, $r_{i,t}$; these continue to be generated via the standard collaborative filtering algorithm that provides the recommendations on the platform. 
Thus, we want our procedure to elicit beliefs both before and after recommendations, so that we can understand how beliefs evolve both with and without recommendation. 
Due to this, we will ensure to choose some subset of movies from $r_{i,t}$ to show up in $B_i^t$.

\raggedbottom

\begin{figure*}[ht]
    \caption{Data Collection Interface}
    \label{fig:belief_elicitation_interface}
    \includegraphics[scale=0.25]{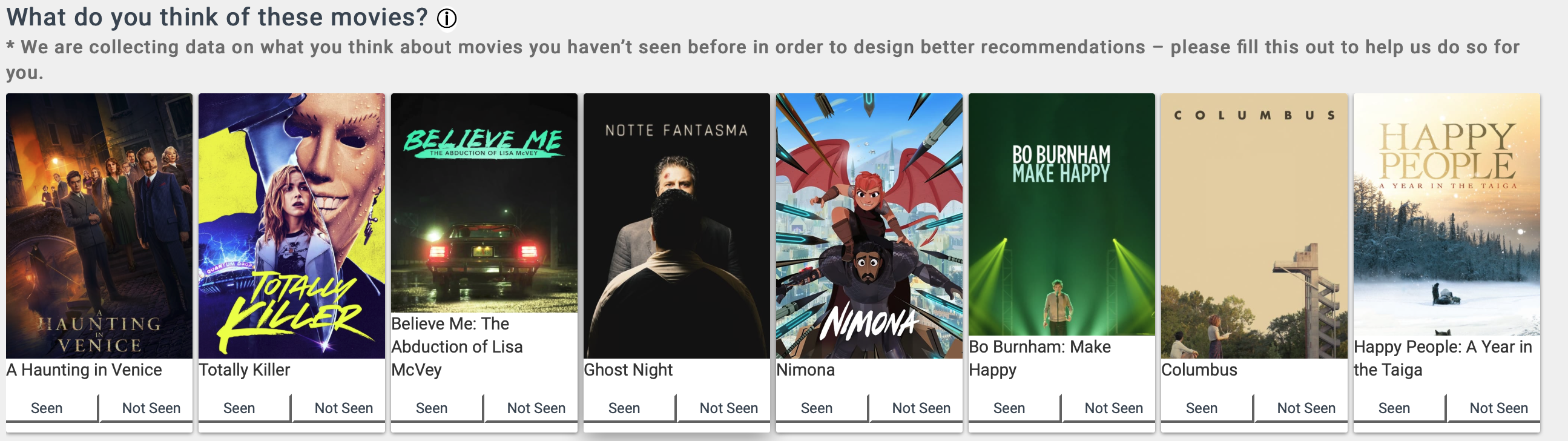}
    \includegraphics[scale=0.4]{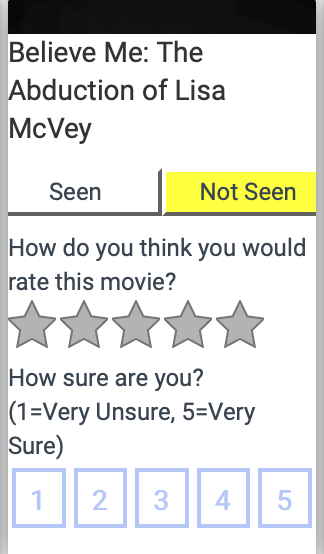}
    \Description[A screenshot of the platform interface]{This is a screenshot of the row of 8 movies we elicit beliefs about. It shows that users see the movie title and can mark seen/unseen. The second panel shows that if they click not seen then it asks them what they think they would rate and how sure they are of this using a Likert scale.}
\end{figure*}

\subsection{Choosing the Set of Goods, $M^t$}

We discuss the procedure that we use to choose $M^{t}$, which is common to all users and refreshed at the beginning of every month. 
In order to ensure that we are getting a wide coverage of the set of possible movies, we design our sampling procedure based on movie genres.\footnote{
    There are 18 genres defined on the platform:
    Action, Adventure, Animation, Comedy, Crime, Documentary, Drama, Fantasy, History, Horror, Music, Mystery, Romance, Science Fiction, TV Movie, Thriller, War, Western.
} We denote the set of genres by $G$.
For each genre $g \in G$, define the fraction of movies that are listed under that genre, $s_g := $ (number of movies of that genre/ total number of movies).\footnote{
    Genres do not partition the set of all movies --- the same movie may be listed under multiple genres --- and therefore we expect $\sum_{g \in G} s_g>1$.
    This will be dealt with by removing duplicates.
} We construct $M^t$ by including movies according to different criteria.
The exact size of $M^t$ is calibrated by a parameter $y$.
The criteria for inclusion are the following:
\begin{enumerate}[label=(\roman*),topsep=.5em,]
    \item \textbf{All-time popularity:} 
    the $50 y$ most popular movies within the different genres $G$. 
    Popularity is determined by the number of ratings.
    For each genre $g$, the $\text{ceil}(s_g \cdot 50 y)$ movies with the most ratings within the genre.
    \item \textbf{Rating:} the $25 y$ movies with the highest rating score within the different genres $G$.\footnote{
        The rating score of the movie is given by the product of its percentile in number of ratings and its percentile in average rating.
    }
    For each genre $g$, select the $\text{ceil}(s_g \cdot 25 y)$ highest rating score within the genre.
    \item  \textbf{Popular recent releases:} the $10 y$ most popular, recently released movies within the different genres $G$. 
    For each genre $g$, select the $\text{ceil}(s_g \cdot 10 y)$ most rated movies that were released within the genre.\footnote{
        A recently released movie is one that was released within the previous recent\_threshold months, where we set $recent\_threshold = 6$.
    }
    \item \textbf{Trendy releases:} the $10 y$ `trendiest' movies within the different genres $G$. 
    For each genre $g$, select $\text{ceil}(s_g \cdot 10 y)$ movies with the highest trendy score within the genre.\footnote{
        The trendy score of a movie is 0 if it has fewer than $num\_rating\_threshold$ ratings or if it has had a non-positive change in ratings since the previous month; otherwise, it is given by (number of ratings today - number of ratings 1 month earlier) $\cdot $ $\ln$(number of ratings today - number of ratings 1 month earlier)/ number of ratings today. 
        We set num\_rating\_threshold = 100.
    }
    \item \textbf{Serendipity:} $5 y$ movies uniformly sampled within the different genres $G$. 
    For each genre $g$, select $\text{ceil}(s_g \cdot 5 y)$ sampled uniformly at random within the genre. 
\end{enumerate}
This leads to having $M^t$ with approximately $100 y$ movies. While we have uniform sampling within each genre in the product space, so that there is a non-zero chance of selecting niche goods, the procedure has a bias towards more popular movies and explicitly accounts for the fact that the set of movies changes over time.

\subsection{Choosing Elicited Goods, $B_i^t$}
At any given period, we present 8 movies to user $i$ to elicit beliefs about, $B_i^t$, chosen from  $M^t$. This appears as a row on the platform's homepage. 
If a user refreshes the page, it generates a new set $B_i^t$, replacing the movies for which beliefs were already elicited.

We do not sample from the full set of $M^{t}$ as the user may have already rated some subset of the movies and we only elicit beliefs about movies that users \textit{have not} seen. 
As such, for a given user denote $M_{i}^{t}\subseteq M^{t}$ as the subset of these that user $i$ has not rated by time $t$ and $R_i^t\subseteq M_i^t$ the further subset for which there are predicted ratings for user $i$ at time $t$.\footnote{
    Note that $B_i^t \subseteq M_i^t$: there may be movies in $R_i^t$ (with predicted rating) and others without any predicted rating. 
    This is so that (1) we do not restrict ourselves to movies with predicted rating, nor (2) require predicted ratings to be computed.
}
 
We choose the set of movies to elicit beliefs about for a user at a given time period $t$, $B_i^t$, according to the following principles: 
\begin{enumerate}[label=-,labelindent=.0em, itemindent=-0.0em, leftmargin=1.0em,topsep=.5em]
    \item 
        \textbf{Broad Sampling}: 3 movies from $M_i^t$ uniformly at random.
    \item 
        \textbf{Elicitation with possible recommendation}: 4 movies from $R_i^t$ sampled uniformly at random from the top $n$ according to their top picks (where $n = 100$).
        This serves the goal of generating variation in the set of elicited movies that may also possibly be recommended, while also sampling from more niche movies due to individual-level heterogeneity in user recommendations.
    \item 
    \textbf{Sample New Movies}: 1 movie from $M_i^t$ that is a recent release.
\end{enumerate}

We impose the additional restriction that we do not elicit a movie if it was elicited two times in the past three months to avoid recurrently eliciting beliefs over the same set of movies.

\section{Data}\label{sec:data}

We first overview the details of the data collection procedure and then provide detailed descriptions of the datasets that we release in order to aide researchers in using them.

\subsection{Data Collection Details}

We provide an overview of the data collection details and process. 
We collect data on the MovieLens platform \citep{HarperKonstan2015ACMTIIS}, whose ratings data constitute a benchmark dataset in the recommender system community. 
We still collect the ratings data and users observe the set of recommendations (``top picks") in the first row on the platform. Users can input ratings directly on the homepage or via searching for a particular movie. 
We do not modify these during the intervention nor do we modify any of the existing rows. 
The only change is that our intervention changes the second row of the homepage of the platform to ask users about the movies chosen by the procedure. We do this so that the data collection is non-invasive and blended into the platform in a non-intrusive manner as our goal is to collect a constant stream of data without over-taxing users' attention. An alternative approach would be to make the data collection a separate survey, as in \cite{aridor2023economics}, but this is much more taxing for users to continually complete and may lead to selection in which users provide this information over long periods of time.

~\autoref{fig:belief_elicitation_interface} provides screenshots of the interface that users see. 
Since this is the only portion of the website where we ask about beliefs data we explicitly provide a description clarifying that we care about understanding their thoughts about movies that they have not seen before. 
Similar to the interface in the surveys in \cite{aridor2023economics}, users can explicitly mark whether they have seen or not seen the movie. 
The right panel of ~\autoref{fig:belief_elicitation_interface} displays the interface observed by the user upon clicking not seen. 
The user inputs their expected rating for a movies on a Likert scale identical to the one used for ratings (from 0.5 to 5, in 0.5 point increments)
Additionally, they also declare how sure they are of their assessment, on a Likert scale from 1 to 5, in 1 point increments. 
If the user clicks that they have seen the movie then they are asked to input a rating and an approximate watch date.

The data collection began at the start of March 2023 and concluded in May 2024 . 
We began the data collection by setting $y = 100$ for $M_{t}$, but revised this down to $y = 11$ in July 2023 once we had a better sense of users' response rates.

\subsection{Datasets}\label{subsec:datasets}

We make publicly available the newly collected beliefs data at \url{https://grouplens.org/datasets/movielens/ml_belief_2024/}. We log both when a user inputs their beliefs about a movie as well as every time a user is prompted to input a belief. 
This allows us to measure selection into elicitation, which may be useful for researchers incorporating this data going forward. 
The dataset contains the following fields, beyond the timestamp and user/movie identifiers:
\begin{enumerate}[label=-,labelindent=.0em, itemindent=-0.0em, leftmargin=1.0em,topsep=.5em]
    \item isSeen: This is $-1$ if the user did not respond, $0$ if the user marked that they had not seen the movie, and $1$ if the user marked that they had seen the movie.
    \item userElicitRating / watchDate: This is the rating the user gave for the movie and the time they claim to have seen it.
    \item userPredictRating / userCertainty: This is the predicted rating and certainty level the user has conditional on them not having seen the movie.
\end{enumerate}
In addition, we provide ratings data that has a similar structure to the main MovieLens ratings datasets that have historically been released by GroupLens, except that the set of users are those that provided at least one belief data point.
Beyond this, we also make available the full logs of recommendations that were presented to these users. 
Combined with the variation that we generate in elicitations, this will allow researchers to explore how recommendations impact beliefs as well as possibly consumption choices.

\section{Descriptive Patterns in the Data}

We now document descriptive patterns in the data that document the extent to which the data has the volume and variation to possibly be suitable for producing recommendations following the proposal in~\autoref{subsec:econ_model}. 
Our goal is to examine the following issues:
\begin{enumerate}[label=-,labelindent=.0em, itemindent=-0.0em, leftmargin=1.0em,topsep=.5em]
    \item \textbf{Response Selection and Volume}: Is there selection in which types of movies users provide their beliefs about? Is there heterogeneity across individuals? Furthermore, what is the volume and distribution of beliefs across users and movies?
    \item \textbf{Belief Validity}: Are users providing meaningful responses (i.e., internally consistent and consistent with reasonable expectations about the data)? 
    \item \textbf{Recommendation-Belief Variation}: How much variation in beliefs and recommendations does the procedure generate? In particular, how many times do movies that we elicit beliefs about also show up in their recommendations?
\end{enumerate}

\noindent \textbf{Response Selection and Volume}: 
We explore whether there is user- or movie-level heterogeneity in response rates. 
Heterogeneity in response rates has been documented for ratings data and that it can be modulated by social cues \citep{chen2010social}.

We first look at the overall volume of belief elicitations. 
We observe 3,004,020 elicitation requests with 28,457 belief responses for 7,518 distinct movies and from 3,199 distinct users. 
19,182 users never provide a response to the belief elicitations. 
Conditional on providing at least one belief response, the response ratio (fraction of requests with a belief response) distribution per user has a mean of 0.078 and a median of 0.031. Furthermore, on average, each user provided beliefs about 8.52 movies. These results indicate that, similar to traditional ratings data, the distribution of belief responses is skewed. 
We now explore whether there is movie-level heterogeneity in response rates by regressing response rate on popularity ($-0.0002, p > 0.1$) and the empirical variance of community ratings ($0.0001, p > 0.1$). 
The results show that users are not more likely to provide beliefs for less popular movies nor to movies with a higher empirical variance of community ratings. Furthermore, the $R^2$ is approximately zero, indicating that these both have little explanatory power for response rate and thus we conclude there is little movie-level selection.

Apart from the novel data on beliefs that we provide, we additionally include the traditional ratings data for the set of users that have provided at least one belief elicitation. We provide the full set of historical ratings for these users, not only those during the intervention period, which contains 1,723,753 total ratings. This makes the ratings data comparable to the MovieLens-1M data in size, but additionally includes the rich data on beliefs and the full set of presented recommendations to these users during the data collection period. As such, while there has been work that explores the value of auxiliary information about the characteristics of movies (e.g., \citep{eden2022investigating, gaag2015froy}) and combining this with MovieLens ratings data, the belief data is complementary and conceptually distinct from these as a combination of these datasets provides a picture of what users think about goods as well as how this varies across the product space.

Overall, these statistics indicate that the scale of the beliefs dataset we have collected on the MovieLens platform can be used for prototyping algorithms that incorporate such data. 
The most natural direction forward would be to apply collaborative filtering and content-based to tackle challenge (1) -- characterizing the full set of beliefs over the product space. 
Finally, while the beliefs data is not at a similar scale as the traditional MovieLens dataset, the response rates and patterns of selection indicate that the procedure implemented on a larger-scale platform or for a sufficiently long time on the MovieLens platform would lead to a reasonably sized dataset that could be used in production recommender systems.

\vspace*{.5em}
\noindent \textbf{Belief Validity}: 
We showcase that the belief data exhibits reasonable patterns. 
We replicate parts of the validation exercises on the belief data conducted by \cite{aridor2023economics} and show that, despite the differing data generation process, similar belief patterns hold. 
For instance, regressing user uncertainty on the log of the total number of ratings has a coefficient of $-0.011$ ($p < 0.001$) which indicates that users are surer of what they think about popular movies. 
Furthermore, estimating a linear probability model for whether a movie was watched on its expected quality ($0.028, p < 0.001)$ and user uncertainty ($-0.134, p < 0.001$) leads to quantitatively similar estimates as \cite{aridor2023economics}: users are more likely to watch movies they expect to like more and are less uncertain about. 
Thus, the different data generation procedure on the platform leads to similar behavioral patterns. 
Researchers can use ratings history data and the rich movie details available on the MovieLens platform to augment the beliefs data and provide better predictions of how they map to consumption.

\vspace*{.5em}
\noindent \textbf{Recommendation-Belief Variation}: 
We explore the extent to which the data has the variation needed to overcome challenge (3) -- variation of both belief elicitations mixed with recommendations. 
We find that, for the average user, 5.9\% of the movies with elicitations requested also show up in their recommendations at some point in the sample. 
Furthermore, when we look at the resulting set of collected beliefs, 9.5\% of the recorded beliefs for the average user also show up in recommendations. Of the three challenges, the data collected from the procedure is most limited in identifying how recommendations shift beliefs.
Nevertheless, this issue can be addressed by deploying the procedure more broadly to obtain more data and adjusting it in line with the experimental results from \cite{aridor2023economics}.

\section{Conclusion}

In conclusion, we have argued that the collection of data on pre-choice attitudes can be important for both designing good recommendations and evaluating their success. One challenge with the collection of such data is that it is not naturally generated from usage of the platform. Motivated by this, we have designed a procedure that chooses which subset of goods to elicit beliefs about in such a way that it is feasible to get reasonable coverage of the product space. We have implemented this procedure and collected data using it for over a year on the MovieLens platform. The resulting dataset is open-sourced and publicly available. We believe that both the procedure and the resulting dataset will help guide the next generation of recommender systems.

\vspace{0.3cm}

\noindent \textbf{Acknowledgements.} We thank Andrew D'Amico for excellent research assistance for this project.

\bibliographystyle{ACM-Reference-Format}
\bibliography{refs}


\begin{thebibliography}{19}


\ifx \showCODEN    \undefined \def \showCODEN     #1{\unskip}     \fi
\ifx \showDOI      \undefined \def \showDOI       #1{#1}\fi
\ifx \showISBNx    \undefined \def \showISBNx     #1{\unskip}     \fi
\ifx \showISBNxiii \undefined \def \showISBNxiii  #1{\unskip}     \fi
\ifx \showISSN     \undefined \def \showISSN      #1{\unskip}     \fi
\ifx \showLCCN     \undefined \def \showLCCN      #1{\unskip}     \fi
\ifx \shownote     \undefined \def \shownote      #1{#1}          \fi
\ifx \showarticletitle \undefined \def \showarticletitle #1{#1}   \fi
\ifx \showURL      \undefined \def \showURL       {\relax}        \fi
\providecommand\bibfield[2]{#2}
\providecommand\bibinfo[2]{#2}
\providecommand\natexlab[1]{#1}
\providecommand\showeprint[2][]{arXiv:#2}

\bibitem[Abdollahpouri et~al\mbox{.}(2020)]%
        {abdollahpouri2020connection}
\bibfield{author}{\bibinfo{person}{Himan Abdollahpouri},
  \bibinfo{person}{Masoud Mansoury}, \bibinfo{person}{Robin Burke}, {and}
  \bibinfo{person}{Bamshad Mobasher}.} \bibinfo{year}{2020}\natexlab{}.
\newblock \showarticletitle{The connection between popularity bias,
  calibration, and fairness in recommendation}. In
  \bibinfo{booktitle}{\emph{Proceedings of the 14th ACM conference on
  recommender systems}}. \bibinfo{pages}{726--731}.
\newblock


\bibitem[Adomavicius and Tuzhilin(2005)]%
        {adomavicius2005toward}
\bibfield{author}{\bibinfo{person}{Gediminas Adomavicius} {and}
  \bibinfo{person}{Alexander Tuzhilin}.} \bibinfo{year}{2005}\natexlab{}.
\newblock \showarticletitle{Toward the next generation of recommender systems:
  A survey of the state-of-the-art and possible extensions}.
\newblock \bibinfo{journal}{\emph{IEEE Transactions on Knowledge \& Data
  Engineering}} \bibinfo{number}{6} (\bibinfo{year}{2005}),
  \bibinfo{pages}{734--749}.
\newblock


\bibitem[Aridor et~al\mbox{.}(2023)]%
        {aridor2023economics}
\bibfield{author}{\bibinfo{person}{Guy Aridor}, \bibinfo{person}{Duarte
  Goncalves}, \bibinfo{person}{Daniel Kluver}, \bibinfo{person}{Ruoyan Kong},
  {and} \bibinfo{person}{Joseph Konstan}.} \bibinfo{year}{2023}\natexlab{}.
\newblock \showarticletitle{The Economics of Recommender Systems: Evidence from
  a Field Experiment on MovieLens}. In \bibinfo{booktitle}{\emph{Proceedings of
  the 24th ACM Conference on Economics and Computation}} (London, United
  Kingdom) \emph{(\bibinfo{series}{EC '23})}. \bibinfo{publisher}{Association
  for Computing Machinery}, \bibinfo{address}{New York, NY, USA},
  \bibinfo{pages}{117}.
\newblock
\showISBNx{9798400701047}
\urldef\tempurl%
\url{https://doi.org/10.1145/3580507.3597677}
\showDOI{\tempurl}


\bibitem[Aridor et~al\mbox{.}(2020)]%
        {AridorGoncalvesSikdar2020ACMRecSys}
\bibfield{author}{\bibinfo{person}{Guy Aridor}, \bibinfo{person}{Duarte
  Gon\c{c}alves}, {and} \bibinfo{person}{Shan Sikdar}.}
  \bibinfo{year}{2020}\natexlab{}.
\newblock \showarticletitle{Deconstructing the Filter Bubble: User
  Decision-Making and Recommender Systems}.
\newblock \bibinfo{journal}{\emph{Fourteenth ACM Conference on Recommender
  Systems}} (\bibinfo{year}{2020}), \bibinfo{pages}{82--91}.
\newblock


\bibitem[Castells et~al\mbox{.}(2021)]%
        {castells2021novelty}
\bibfield{author}{\bibinfo{person}{Pablo Castells}, \bibinfo{person}{Neil
  Hurley}, {and} \bibinfo{person}{Saul Vargas}.}
  \bibinfo{year}{2021}\natexlab{}.
\newblock \showarticletitle{Novelty and diversity in recommender systems}.
\newblock In \bibinfo{booktitle}{\emph{Recommender systems handbook}}.
  \bibinfo{publisher}{Springer}, \bibinfo{pages}{603--646}.
\newblock


\bibitem[Chen et~al\mbox{.}(2010)]%
        {chen2010social}
\bibfield{author}{\bibinfo{person}{Yan Chen}, \bibinfo{person}{F~Maxwell
  Harper}, \bibinfo{person}{Joseph Konstan}, {and} \bibinfo{person}{Sherry~Xin
  Li}.} \bibinfo{year}{2010}\natexlab{}.
\newblock \showarticletitle{Social comparisons and contributions to online
  communities: A field experiment on movielens}.
\newblock \bibinfo{journal}{\emph{American Economic Review}}
  \bibinfo{volume}{100}, \bibinfo{number}{4} (\bibinfo{year}{2010}),
  \bibinfo{pages}{1358--98}.
\newblock


\bibitem[Eden et~al\mbox{.}(2022)]%
        {eden2022investigating}
\bibfield{author}{\bibinfo{person}{Sagi Eden}, \bibinfo{person}{Amit Livne},
  \bibinfo{person}{Oren Sar~Shalom}, \bibinfo{person}{Bracha Shapira}, {and}
  \bibinfo{person}{Dietmar Jannach}.} \bibinfo{year}{2022}\natexlab{}.
\newblock \showarticletitle{Investigating the Value of Subtitles for Improved
  Movie Recommendations}. In \bibinfo{booktitle}{\emph{Proceedings of the 30th
  ACM Conference on User Modeling, Adaptation and Personalization}}.
  \bibinfo{pages}{99--109}.
\newblock


\bibitem[Gaag et~al\mbox{.}(2015)]%
        {gaag2015froy}
\bibfield{author}{\bibinfo{person}{Philip Gaag}, \bibinfo{person}{Daniel
  Granvogl}, \bibinfo{person}{Robert Jackermeier}, \bibinfo{person}{Florian
  Ludwig}, \bibinfo{person}{Johannes Rosenl{\"o}hner}, {and}
  \bibinfo{person}{Alexander Uitz}.} \bibinfo{year}{2015}\natexlab{}.
\newblock \showarticletitle{FROY: exploring sentiment-based movie
  recommendations}. In \bibinfo{booktitle}{\emph{Proceedings of the 14th
  International Conference on Mobile and Ubiquitous Multimedia}}.
  \bibinfo{pages}{345--349}.
\newblock


\bibitem[Ge et~al\mbox{.}(2010)]%
        {ge2010beyond}
\bibfield{author}{\bibinfo{person}{Mouzhi Ge}, \bibinfo{person}{Carla
  Delgado-Battenfeld}, {and} \bibinfo{person}{Dietmar Jannach}.}
  \bibinfo{year}{2010}\natexlab{}.
\newblock \showarticletitle{Beyond accuracy: evaluating recommender systems by
  coverage and serendipity}.
\newblock \bibinfo{journal}{\emph{Proceedings of the fourth ACM conference on
  Recommender systems}} (\bibinfo{year}{2010}), \bibinfo{pages}{257--260}.
\newblock


\bibitem[Harper and Konstan(2015)]%
        {HarperKonstan2015ACMTIIS}
\bibfield{author}{\bibinfo{person}{F.~Maxwell Harper} {and}
  \bibinfo{person}{Joseph~A. Konstan}.} \bibinfo{year}{2015}\natexlab{}.
\newblock \showarticletitle{The MovieLens datasets: History and context}.
\newblock \bibinfo{journal}{\emph{ACM Transactions on Interactive Intelligent
  Systems (TIIS)}} \bibinfo{volume}{5}, \bibinfo{number}{4}
  (\bibinfo{year}{2015}), \bibinfo{pages}{1--19}.
\newblock


\bibitem[Kaminskas and Bridge(2016)]%
        {kaminskas2016diversity}
\bibfield{author}{\bibinfo{person}{Marius Kaminskas} {and}
  \bibinfo{person}{Derek Bridge}.} \bibinfo{year}{2016}\natexlab{}.
\newblock \showarticletitle{Diversity, serendipity, novelty, and coverage: a
  survey and empirical analysis of beyond-accuracy objectives in recommender
  systems}.
\newblock \bibinfo{journal}{\emph{ACM Transactions on Interactive Intelligent
  Systems (TiiS)}} \bibinfo{volume}{7}, \bibinfo{number}{1}
  (\bibinfo{year}{2016}), \bibinfo{pages}{1--42}.
\newblock


\bibitem[Kotkov et~al\mbox{.}(2018)]%
        {kotkov2018investigating}
\bibfield{author}{\bibinfo{person}{Denis Kotkov}, \bibinfo{person}{Joseph~A
  Konstan}, \bibinfo{person}{Qian Zhao}, {and} \bibinfo{person}{Jari
  Veijalainen}.} \bibinfo{year}{2018}\natexlab{}.
\newblock \showarticletitle{Investigating serendipity in recommender systems
  based on real user feedback}. In \bibinfo{booktitle}{\emph{Proceedings of the
  33rd annual acm symposium on applied computing}}.
  \bibinfo{pages}{1341--1350}.
\newblock


\bibitem[Kotkov et~al\mbox{.}(2016)]%
        {kotkov2016challenges}
\bibfield{author}{\bibinfo{person}{Denis Kotkov}, \bibinfo{person}{Jari
  Veijalainen}, {and} \bibinfo{person}{Shuaiqiang Wang}.}
  \bibinfo{year}{2016}\natexlab{}.
\newblock \showarticletitle{Challenges of serendipity in recommender systems}.
\newblock \bibinfo{journal}{\emph{WEBIST 2016: Proceedings of the 12th
  International conference on web information systems and technologies. Volume
  2, ISBN 978-989-758-186-1}} (\bibinfo{year}{2016}).
\newblock


\bibitem[McNee et~al\mbox{.}(2006)]%
        {mcnee2006being}
\bibfield{author}{\bibinfo{person}{Sean~M McNee}, \bibinfo{person}{John Riedl},
  {and} \bibinfo{person}{Joseph~A Konstan}.} \bibinfo{year}{2006}\natexlab{}.
\newblock \showarticletitle{Being accurate is not enough: how accuracy metrics
  have hurt recommender systems}.
\newblock \bibinfo{journal}{\emph{CHI'06 extended abstracts on Human factors in
  computing systems}} (\bibinfo{year}{2006}), \bibinfo{pages}{1097--1101}.
\newblock


\bibitem[Nguyen et~al\mbox{.}(2014)]%
        {NguyenHuiHarperTerveenKonstan2014ACM}
\bibfield{author}{\bibinfo{person}{Tien~T. Nguyen}, \bibinfo{person}{Pik-Mai
  Hui}, \bibinfo{person}{F.~Maxwell Harper}, \bibinfo{person}{Loren Terveen},
  {and} \bibinfo{person}{Joseph~A. Konstan}.} \bibinfo{year}{2014}\natexlab{}.
\newblock \showarticletitle{Exploring the filter bubble: the effect of using
  recommender systems on content diversity}.
\newblock \bibinfo{journal}{\emph{Proceedings of the 23rd International
  Conference on World Wide Web}} (\bibinfo{year}{2014}),
  \bibinfo{pages}{677--686}.
\newblock


\bibitem[Schnabel et~al\mbox{.}(2016)]%
        {schnabel2016recommendations}
\bibfield{author}{\bibinfo{person}{Tobias Schnabel}, \bibinfo{person}{Adith
  Swaminathan}, \bibinfo{person}{Ashudeep Singh}, \bibinfo{person}{Navin
  Chandak}, {and} \bibinfo{person}{Thorsten Joachims}.}
  \bibinfo{year}{2016}\natexlab{}.
\newblock \showarticletitle{Recommendations as treatments: Debiasing learning
  and evaluation}.
\newblock \bibinfo{journal}{\emph{Proceedings of the International Conference
  on Machine Learning}} (\bibinfo{year}{2016}), \bibinfo{pages}{1670--1679}.
\newblock


\bibitem[Steck(2018)]%
        {steck2018calibrated}
\bibfield{author}{\bibinfo{person}{Harald Steck}.}
  \bibinfo{year}{2018}\natexlab{}.
\newblock \showarticletitle{Calibrated recommendations}.
\newblock \bibinfo{journal}{\emph{Proceedings of the 12th ACM conference on
  recommender systems}} (\bibinfo{year}{2018}), \bibinfo{pages}{154--162}.
\newblock


\bibitem[Vargas and Castells(2011)]%
        {vargas2011rank}
\bibfield{author}{\bibinfo{person}{Sa{\'u}l Vargas} {and}
  \bibinfo{person}{Pablo Castells}.} \bibinfo{year}{2011}\natexlab{}.
\newblock \showarticletitle{Rank and relevance in novelty and diversity metrics
  for recommender systems}.
\newblock \bibinfo{journal}{\emph{Proceedings of the fifth ACM conference on
  Recommender systems}} (\bibinfo{year}{2011}), \bibinfo{pages}{109--116}.
\newblock


\bibitem[Ying et~al\mbox{.}(2006)]%
        {ying2006leveraging}
\bibfield{author}{\bibinfo{person}{Yuanping Ying}, \bibinfo{person}{Fred
  Feinberg}, {and} \bibinfo{person}{Michel Wedel}.}
  \bibinfo{year}{2006}\natexlab{}.
\newblock \showarticletitle{Leveraging missing ratings to improve online
  recommendation systems}.
\newblock \bibinfo{journal}{\emph{Journal of marketing research}}
  \bibinfo{volume}{43}, \bibinfo{number}{3} (\bibinfo{year}{2006}),
  \bibinfo{pages}{355--365}.
\newblock


\end{thebibliography}


\end{document}